\pdfoutput=1
\RequirePackage{ifpdf}
\ifpdf % We~are running pdfTeX in pdf mode
\documentclass[pdftex]{sigma}
\else
\documentclass{sigma}
\fi

\usepackage{mathtools}

\numberwithin{equation}{section}

\newtheorem*{Theorem*}{Theorem}

 { \theoremstyle{definition}

 }

\newcommand{\hitein}{h }
\newcommand{\hito}{g }
\newcommand{\CV}{\mathcal{V}}
\newcommand{\CVe}{\mathcal{V}_ \mathrm{E} }
\newcommand{\stare}{\star_{ \mathrm{E} }}
\newcommand{\CVa}{\CV_{ \mathrm{a} }}
\newcommand{\omegae}{\omega _{ \mathrm{E} }}
\newcommand{\omegaa}{\omega _{ \mathrm{a} }}
\newcommand{\tu}[1]{\textup{#1}}
\newcommand{\rd}{\mathrm{d}}

\DeclareMathOperator{\arccosh}{arcCosh}

\newcommand{\lred}{L_{\tu{red}}}

\newcommand{\red}[1]{}
\newcommand{\blue}[1]{}
\newcommand{\ora}[1]{{#1 }}

\newcommand{\gK}{g _K}
\newcommand{\rad}{V}
\newcommand{\La}{L_{ \mathrm{a} }}
\newcommand{\ghtn}{g _{ \mathrm{hTN} }}

\begin{document}
\allowdisplaybreaks

\renewcommand{\thefootnote}{}

\newcommand{\arXivNumber}{2302.13792}

\renewcommand{\PaperNumber}{043}

\FirstPageHeading

\ShortArticleName{The Asymptotic Structure of the Centred Hyperbolic 2-Monopole Moduli Space}

\ArticleName{The Asymptotic Structure of the Centred Hyperbolic 2-Monopole Moduli Space\footnote{This paper is a~contribution to the Special Issue on Topological Solitons as Particles. The~full collection is available at \href{https://www.emis.de/journals/SIGMA/topological-solitons.html}{https://www.emis.de/journals/SIGMA/topological-solitons.html}}}

\Author{Guido FRANCHETTI~$^{\rm a}$ and Calum ROSS~$^{\rm b}$}

\AuthorNameForHeading{G.~Franchetti and C.~Ross}

\Address{$^{\rm a)}$ Department of Mathematical Sciences, University of Bath,\\
\hphantom{$^{\rm a)}$}~Claverton Down, Bath BA2 7AY, England, UK}
\EmailD{\href{mailto:gf424@bath.ac.uk}{gf424@bath.ac.uk}}

\Address{$^{\rm b)}$~Department of Mathematics, University College London,\\
\hphantom{$^{\rm b)}$}~London, WC1E 6BT, England, UK}
\EmailD{\href{mailto:calum.ross@ucl.ac.uk}{calum.ross@ucl.ac.uk}}

\ArticleDates{Received February 28, 2023, in final form June 21, 2023; Published online July 04, 2023}

\Abstract{We construct an asymptotic metric on the moduli space of two centred hyperbolic monopoles by working in the point particle approximation, that is treating well-separated monopoles as point particles with an electric, magnetic and scalar charge and re-interpreting the dynamics of the 2-particle system as geodesic motion with respect to some metric. The corresponding analysis in the Euclidean case famously yields the negative mass Taub-NUT metric, which asymptotically approximates the $L ^2 $ metric on the moduli space of two Euclidean monopoles, the Atiyah--Hitchin metric. An important difference with the Euclidean case is that, due to the absence of Galilean symmetry, in the hyperbolic case it is not possible to factor out the centre of mass motion. Nevertheless we show that we can consistently restrict to a 3-dimensional configuration space by considering antipodal configurations. In complete parallel with the Euclidean case, the metric that we obtain is then the hyperbolic analogue of negative mass Taub-NUT. We also show how the metric obtained is related to the asymptotic form of a hyperbolic analogue of the Atiyah--Hitchin metric constructed by Hitchin.}


\Classification{70S15; 14D21}

\begin{flushright}
\begin{minipage}{63mm}
\it To Nick Manton on his 70th birthday
\end{minipage}
\end{flushright}

\renewcommand{\thefootnote}{\arabic{footnote}}
\setcounter{footnote}{0}

\section{Introduction}
\label{introduction}
Magnetic monopoles~\cite{manton:2004} are an interesting class of topological solitons defined on a Riemannian 3-manifold $M$. The monopole data consists of a pair $(A, \Phi )$, where $A$ is a connection on a~principal ${\rm SU}(2) $-bundle over $M$ and $\Phi $ is a section of the associated adjoint bundle. The pair $( A , \Phi ) $ satisfies a system of first order PDEs known as the Bogomolny equations supplemented by suitable boundary conditions. In order for the Bogomolny equations to admit non-singular solutions $M$ must be non-compact; the cases of Euclidean 3-space $E ^3 $ and hyperbolic 3-space~$H ^3 $ have received the most attention.

Hyperbolic and Euclidean monopoles share many similarities. For example, in both cases the space of solutions of the Bogomolny equations is a smooth manifold of dimension $4|k| $, where $k$ is a topological integer which counts the total magnetic charge of the monopole solution. At least for well-separated configurations, $| k |$ can be interpreted as the number of monopoles described by the solution.

There are however a number of important differences between the two cases as we now discuss.
First, for the class of boundary conditions usually considered, the Higgs field norm $\| \Phi \| $ of both Euclidean and hyperbolic monopoles has a finite non-zero limit, known as the monopole mass $p$, as we move to infinity which is independent of the direction. More precisely, both $E ^3 $ and $H ^3 $ admit a cohomogeneity one action of ${\rm SO}(3) $ with $S ^2 $ as the typical orbit. Let $r$ be a~coordinate transverse to the ${\rm SO}(3) $ orbits such that the sphere volume increases with $r$. Then $p =R   \lim _{ r \rightarrow \infty }\| \Phi \| $. In the Euclidean case $p$ can always be fixed to any non-zero positive value, traditionally one, by rescaling. However, due to the length scale $R$ associated to the non-zero curvature $- R ^{ -2 }$ of hyperbolic space, the mass of a hyperbolic monopoles cannot be fixed and $p $ is an effective parameter.

It is worth noting that monopoles such that $2p \in \mathbb{Z} $ are equivalent to circle-invariant instantons on $E^4 $~\cite{atiyah:1988}. The monopole number $k$ and the instanton number $I$ are related by $I =2k p $. While Euclidean monopoles could be similarly related to translation invariant instantons on $E ^4 $, translation invariance would cause the instanton to have infinite action and the correspondence becomes much less useful.

Second, in both cases the monopole abelianises at infinity, i.e., for $r$ the transverse coordinate introduced above, $\Phi |_{ S ^2 _r }$ and $A |_{ S ^2 _r }$ become parallel elements of $\mathfrak{su } (2) $ as $r \rightarrow \infty $. However in the Euclidean case the data induced on $S ^2 _\infty $ only determines the monopole charge $k$, while a~hyperbolic monopole is fully determined by its asymptotic data~\cite{braam:1990}.

The third difference, which constitutes the motivation for this work, has to do with the possibility of equipping the moduli space $\mathcal{M} _k $ of charge $k$ monopoles with a ``natural'' Riemannian metric of physical significance. The (framed) moduli space $\mathcal{M} _k $ is the space of solutions of the Bogomolny equations with the appropriate boundary conditions modulo bundle automorphisms which become the identity at some fixed point of $S ^2_\infty $. For both Euclidean and hyperbolic monopoles it is known to be a smooth manifold of dimension $4k $~\cite{atiyah:1988, donaldson:1984a}. In the Euclidean case, the flat $L ^2 $ metric on the space of field configurations $(A, \Phi ) $ descends to a curved metric on the space of field configurations modulo bundle automorphisms. Restricting to $\mathcal{M} ^{{\rm Eucl}} _k $ yields the $L ^2 $ moduli space metric.

This metric has an important physical interpretation thanks to the adiabatic dynamics approximation: Yang--Mills--Higgs dynamics in $3+1 $ dimensions with initial conditions close to a~solution of the Bogomolny equations is well approximated by geodesic motion on $\mathcal{M} ^{ {\rm Eucl}} _k $ with respect to the $L ^2 $ metric~\cite{manton:1982}. In the case of Euclidean monopoles the $L ^2 $ metric is hyperk\"ahler~\cite{atiyah:2014}. The moduli space $\mathcal{M} ^{ {\rm Eucl}} _k $ has a Riemannian product decomposition $\mathcal{M} ^{ {\rm Eucl}} _k = E ^3 \times \big(S ^1 \times \tilde M _k \big)/ \mathbb{Z} _k $, where $\tilde M _k $ is irreducible simply connected of dimension $4 (k-1 )$. The factor $E ^3 \times S ^1 $ carries the flat product metric. A point in $E ^3 \times S ^1 $ specifies the monopole centre of mass in $E ^3 $ and a phase angle whose time dependence determines the total electric charge. The moduli space metric on~$\tilde M _2 / \mathbb{Z} _2 $ is the celebrated Atiyah--Hitchin metric~\cite{atiyah:2014}.

In the hyperbolic case, the $L ^2 $ metric on $\mathcal{M} ^{ \text{hyp}} _k $ is divergent. Of course, other metrics can be defined and various alternative approaches have been proposed in the literature: the boundary metric originally proposed by Braam--Austin~\cite{braam:1990}, see also~\cite{bolognesi:2014,sutcliffe:2022,sutcliffe:2022a} for further work, the instanton metric restricted to circle-invariant configurations~\cite{franchetti:2017}, the twistorial approach of~\cite{bielawski:2013a,bielawski:2013, nash:2007a} and, for a charge 2 monopole, the family of Einstein metrics constructed in~\cite{hitchin:1996}. The relations between these metrics and their relevance, if any, to the dynamics of magnetic monopoles is still unclear.\looseness=-1

Another difference between the Euclidean and hyperbolic case is that, since there is no analogue of the Galilei group for $\mathbb{R} \times H ^3$, we do not expect the moduli corresponding to the centre of mass position in $H ^3 $ to factorise as they do in the Euclidean case. However, it is still possible to identify an $S^{1}$ factor corresponding to the total electric charge. In other words
$\mathcal{M} ^{ \text{hyp}} _k = \big(S ^1 \times \tilde M ^{ \text{hyp} } _k \big)/ \mathbb{Z} _k
$,
 where $\tilde M ^{ \text{hyp} } _k $ is a simply connected manifold of dimension $4k -1 $.

In this paper we approach the problem starting from point particle dynamics. As shown in~\cite{manton:1985,manton:1985a} for two monopoles and in~\cite{gibbons:1995} for the general case, the asymptotic region of the moduli space of $k$ Euclidean monopoles can be probed by making use of the point particle approximation. That is, well-separated monopoles are approximated by point particles having equal masses and magnetic charges but different electric charges. The resulting $k$-particle dynamics can be re-interpreted as geodesic motion on a $T ^k $-bundle over the configuration space of $k$ particles equipped with a Riemannian metric which can be determined from the Lagrangian of the particle system.

The metric obtained following this procedure is generally incomplete as it develops singularities at finite distances. Since it is only an approximate metric valid in the region of the moduli space corresponding to well-separated monopoles, these singularities are not worrisome and from the physical point of view they just signal the fact that the point particle approximation breaks down as the monopoles come close to each other. For the case of two monopoles, the metric found in~\cite{manton:1985,manton:1985a}, see also~\cite{feher:1987, gibbons:1986}, after fixing the centre of mass of the 2-particle system is the famous Taub-NUT metric~\cite{hawking:1977}. The Taub-NUT metric depends on one effective parameter~$M$ called mass. It is complete for non-negative values of $M$ but becomes singular in the interior if~$M<0$. The metric found in~\cite{manton:1985,manton:1985a} is the negative mass version of Taub-NUT, which is indeed the asymptotic form of the $L ^2 $ metric on $\tilde M _2 / \mathbb{Z} _2 $, the complete Atiyah--Hitchin metric.

Here we carry out the analysis for two particles in $H^3 $. As mentioned, the hyperbolic case is complicated by the fact that $H ^3\times \mathbb{R} $ has no analogue of the Galilei group so in general it is not possible to factor out the centre of mass motion. In fact, in general it is not even clear what the centre of mass should be: even for two particles there are competing definitions which are inequivalent if the particles have different masses~\cite{galperin:1993,garcaa-naranjo:2020,garca-a-naranjo:2016}, and no point satisfies the property of being either fixed or moving along a geodesic for general configurations with pairwise attractive interactions~\cite{diacu:2012,garca-a-naranjo:2016}.

 A general analysis would thus have to consider the full 6-dimensional configuration space of two particles in $H ^3 $. However, it is possible to simplify the problem if we restrict our attention to specific configurations. The isometry group of $H ^3 $ is the (orthochronous subgroup of the) Lorentz group, and acts by symmetries on the particle Lagrangian. The conserved quantities associated to boosts and rotations can be naturally identified with the total linear and angular momenta of the particle system. By the conservation of linear momentum, if the initial conditions are taken so that the two particles are at antipodal positions and have opposite velocities, then the particles will remain antipodal throughout their motion. For such configurations we thus reduce to a 3-dimensional configuration space.

Following the analysis of~\cite{manton:1985,manton:1985a}, we reinterpret particle dynamics restricted to antipodal configurations as geodesic motion on an $S ^1 $ bundle over this 3-dimensional configuration space.
By doing so we obtain a Riemannian metric, which we like to call hyperbolic Taub-NUT~\cite{franchetti:2018} due to its manifest similarities with the Taub-NUT metric, first constructed in~\cite{lebrun:1991a}. The hyperbolic Taub-NUT metric, just like the Taub-NUT one, depends on one effective parameter $M$ called mass, is complete for $M \geq 0 $ and becomes singular in the interior if $M<0 $. In complete analogy with the results of~\cite{manton:1985}, the metric that we obtain is hyperbolic Taub-NUT with negative mass.

It is interesting to note that the metric we obtain corresponds, for $k =2 $, to the one found in~\cite{gibbons:2007} by considering the motion in $H ^3 $ of a monopole in the background of $k-1 $ fixed ones. As already noted in~\cite{gibbons:2007}, while fixing the positions of all but one monopole bypasses the need to deal with a higher dimensional configuration space, it is unphysical from the perspective of~${\rm SU}(2)$ monopoles dynamics since for well separated configurations the mass of each monopole is determined by the other charges and not a free parameter. Therefore, the analysis in~\cite{gibbons:2007} does not allow one to interpret hyperbolic Taub-NUT as a geodesic submanifold of the full moduli-space. Our results instead show that negative mass hyperbolic Taub-NUT does indeed capture the asymptotics of some metric on the moduli space of two centred hyperbolic monopoles.

It is then natural to ask what is the metric which negative mass hyperbolic Taub-NUT is approximating, i.e.,~what is our hyperbolic analogue of the Atiyah--Hitchin manifold. As we discuss in Section~\ref{sec:further remarks}, a metric in the conformal class of the Einstein metric constructed in~\cite{hitchin:1996} asymptotically reduces to hyperbolic Taub-NUT with negative mass, again in complete parallelism with the Euclidean case.

The plan of the paper is as follows: In Section~\ref{sec:point particles}, we discuss our conventions and some useful properties of $H^{3}$, summarise the basics of the point particle approximation, and finally derive a~metric on the asymptotic moduli space of two centred monopoles in $H ^3 $. In Section~\ref{sec:further remarks}, we relate this metric to the hyperbolic analogue of the Atiyah--Hitchin metric constructed by Hitchin in~\cite{hitchin:1996} and discuss some open questions.

\section[Point particle dynamics in H\^{}3]{Point particle dynamics in $\boldsymbol{H^{3}}$}
\label{sec:point particles}
\subsection[Some facts about H\^{}3]{Some facts about $\boldsymbol{H^{3}}$}
Perhaps the most straightforward model of hyperbolic space $H ^3 $ is the ``pseudosphere'' $L$ in Minkowski space $E ^{ 1,3 }$, that is the (upper) hyperboloid
\begin{equation*}
%\label{loidmodel}
L = \big\{ (W,X,Y,Z) \in E ^{1,3} \colon X ^2 + Y ^2 + Z ^2 - W ^2 = - R ^2 , \, W >0 \big\}
\end{equation*}
with the Riemannian metric induced as a submanifold of $E ^{1, 3 }$. The parameter $R$ is related to the curvature $\kappa $ of $H ^3 $ via $\kappa =- R ^{ -2 }$.
Since the constraints defining $L $ are invariant under the subgroup ${\rm O}^+(1,3)$ of the Lorentz group consisting of orthochronous Lorentz transformations, it is clear that $L$ has isometry group ${\rm O}^+(1,3 )$. In these coordinates the Killing vector fields~$(X _i, Y _i )$ generating rotations and boosts have very simple expressions,
\begin{alignat}{4}
\label{kvfLrot}
&X _1 = Y \partial _Z - Z \partial _Y , \qquad&& X _2 = Z \partial _X - X \partial _Z , \qquad&& X _3 = X \partial _Y - Y \partial _X, &\\
\label{kvfLboo}
&Y _1= X \partial _W+W \partial _X , \qquad&& Y _2 = Y \partial _W + W \partial _Y , \qquad&& Y _3 = Z \partial _W + W \partial _Z. &
\end{alignat}
The vector fields~(\ref{kvfLrot}) and~(\ref{kvfLboo}) satisfy the $\mathfrak{so } (1,3 )$ Lie algebra relations,
\begin{equation*}
[ X _i , X _j ] = - \epsilon _{ ijk }X _k , \qquad
[ X _i , Y _j ] = - \epsilon _{ ijk }Y _k , \qquad
[ Y _i ,Y _j ] = + \epsilon _{ ijk }X _k .
\end{equation*}
Geodesics in this model are given by the intersection of $H ^3 $ with 2-planes through the origin. The hyperbolic distance between two points $\mathbf{X}_{1},\mathbf{X}_{2}\in L$, having coordinates $( W _i, X _i , Y _i , Z _i )$, is given by
\begin{equation*}
%\label{hypdistl}
D_L(\mathbf{X}_{1},\mathbf{X}_{2})= R \arccosh\left(-\frac{g_{E ^{1,3}}(\mathbf{X}_{1},\mathbf{X}_{2})}{R^2}\right),
\end{equation*}
where $g _{ E ^{ 1,3 } }$ is the inner product on Minkowski space $E ^{ 1,3 } $.

The Klein--Beltrami model $K$ is obtained by gnomonic projection of $L$: a point $p$ on the hyperboloid is mapped to the intersection point between the straight line (in the Euclidean sense) from $p$ to $(0,0,0,0) \in E ^{1, 3 }$ and the hyperplane $W =R $ tangent to the hyperboloid at~$(R,0,0,0) $. Denoting by $(x,y,z )$ coordinates on $K$, we thus have the relation
\begin{equation}
\label{KtoL}
(x,y,z ) = \frac{R}{W} (X , Y , Z ),
\end{equation}
and we see that
\begin{equation}
\label{kelinm}
K =\big\{ (x,y,z )\in E ^3 \colon x ^2 + y ^2 + z ^2 < R ^2 \big\} ,
\end{equation}
the open ball of radius $R$.
For reference~(\ref{KtoL}) has inverse
\begin{equation}
\label{LtoK}
(X,Y,Z,W) = \frac{ R}{\sqrt{ R ^2 - |\mathbf{x} | ^2 } }( x,y,z,R),
\end{equation}
where $|\mathbf{x} |^2 = x ^2 + y ^2 + z ^2 $. We will often denote by $ \mathbf{x} $ a point in $H ^3 $ having coordinates $(x,y,z )$ in the Klein model $K$. In $K$ the hyperbolic distance between two points $\mathbf{x} _1 $, $\mathbf{x} _2 $ is
\begin{equation}
\label{hypdistk}
D_K(\mathbf{x}_{1},\mathbf{x}_{2})= R \arccosh\left( \frac{R ^2 - g _{ E _3 }(\mathbf{x} _1 , \mathbf{x} _2 )}{\sqrt{R ^2 - |\mathbf{x} _1 |^2 }\sqrt{R ^2 - |\mathbf{x} _2 |^2 }}\right),
\end{equation}
where $g _{ E ^3 }$ is the Euclidean metric on $E ^3 $.
The metric on $K$ is obtained by pulling back that on~$L$ via~(\ref{LtoK}), getting
\begin{equation}
\label{bkmetric}
\gK =R ^2 \Bigg( \frac{\big(R ^2 - |\mathbf{x} |^2 \big) \mathrm{d} \mathbf{x} \cdot \mathrm{d} \mathbf{x} + (\mathbf{x} \cdot \mathrm{d} \mathbf{x} )^2 }{\big(R ^2 - |\mathbf{x} |^2 \big) ^2 }\Bigg) .
\end{equation}

Due to the off-diagonal terms in $\gK $, the Klein--Beltrami model may seem unappealing when compared to other models such as the half-space model or the Poincar\'e one. However it shines in at least two respects. First, all the geodesics in $K$ are straight line segments. Second, the Killing vector fields take a convenient form,
\begin{alignat}{4}
\label{kvfKrot}
&X _1= y \partial _z - z \partial _y , \qquad&& X _2 = z \partial _x - x \partial _z , \qquad&& X _3 = x \partial _y - y \partial _x, & \\
%\label{kvfKboo}
&Y _1= R ^2 \partial _x - x\rad , \qquad&&
Y _2 = R ^2 \partial _y - y \rad , \qquad&&
Y _3 = R ^2 \partial _z - z \rad , & \nonumber
\end{alignat}
where
\begin{equation*}
\rad=x \partial _x + y \partial _y + z \partial _z ,
\end{equation*}
making the interpretation of conserved quantities transparent, cf.~equations~(\ref{consangmom}) and~(\ref{conslinmom}) below. A nice review of the properties of the most common models of hyperbolic space is contained in~\cite{cannon:1997}.

It can be useful to introduce other coordinate systems on $K$. Defining polar coordinates $(\rho , \theta , \phi )$ as in $E ^3 $, with $0 \leq \rho <R $, $\theta \in [0, \pi ] $, $\phi \in [0, 2 \pi )$,
\begin{equation*}
x =\rho \sin \theta \cos \phi , \qquad y =\rho \sin \theta \sin \phi , \qquad z =\rho \cos \theta,
\end{equation*}
(\ref{bkmetric}) becomes
\begin{equation*}
\gK =R ^2 \Bigg( \frac{R ^2 \mathrm{d} \rho ^2 + \big(R ^2 - \rho ^2 \big) \rho ^2 \mathrm{d} \Omega ^2 }{\big(R ^2 - \rho ^2 \big)^2 } \Bigg)
\end{equation*}
for $\rd\Omega ^2 =\mathrm{d} \theta ^2 + \sin ^2 \theta \, \mathrm{d} \phi ^2 $ the round metric on $S ^2 $. If we now redefine the radial variable by
\begin{equation*}
\sinh \left( \frac{r}{R} \right) = \frac{\rho }{\sqrt{ R ^2 - \rho ^2 }}\quad \Leftrightarrow \quad \rho = R \tanh \left( \frac{r}{R} \right)
\end{equation*}
we get
\begin{equation}
\label{bgpolargeo}
\gK =\mathrm{d} r ^2 + R ^2 \sinh ^2 \left( \frac{r}{R} \right) \mathrm{d} \Omega ^2,
\end{equation}
showing that $r \in [0, \infty )$ is a geodesic coordinate.

Spatial inversion $A$ belongs to the isometry group of $H ^3 $, so given any point $p \in H ^3 $ we define its \emph{antipodal} point to be $A (p) $. In the Klein model~(\ref{kelinm}) we simply have $A (x,y,z ) =-(x,y,z) $.

We will need to make use of parallel transport with respect to the Levi-Civita connection~$\nabla $ of $H ^3 $ in order to identify tangent spaces at different points $\mathbf{x}_{1} $, $\mathbf{x}_{2} $. Parallel transport along a~curve $\gamma\colon[0,1] \rightarrow H ^3 $ from $\mathbf{x} _1 $ to $\mathbf{x} _2 $ is an isometry $P ^\gamma _{ 21 }\colon T _{ \mathbf{x}_{1} } H ^3 \rightarrow T _{ \mathbf{x}_{2} } H ^3 $ obtained as follows. Let $v _1 \in T _{ \mathbf{x}_{1} }H ^3 $, then the parallel transport along $\gamma$ of $ v _1 $ is the vector $P ^\gamma _{ 21} v _1 \in T _{ \mathbf{x}_{2} }H ^3 $ obtained evaluating at $t =1 $ the vector field along $\gamma$ which solves the parallel transport ODE with initial condition $V (0) = v _1 $. With respect to a coordinate frame $ \{ \partial _i \} $ the ODE reads
\begin{equation}
\label{pteq}
\frac{\mathrm{d} }{\mathrm{d} t} V ^i + \Gamma ^i _{ j k } U ^j V ^k =0,
\end{equation}
where $\Gamma ^i _{ j k }$ are the Christoffel symbols associated to $\nabla $ and $U ^j $ the components of the vector field tangent to $\gamma$. The inverse of $P _{ 21 } ^\gamma $ is $P _{ 12 }^{ - \gamma }$ where $ (- \gamma) (t )=\gamma (1-t )$ is the same curve with the opposite orientation.

As is well known, parallel transport in a curved space depends on the choice of $\gamma$. An important property of hyperbolic space is that given any two points $\mathbf{x}_{1},\mathbf{x}_{2} \in H ^3 $ there is a~unique length-minimising geodesic connecting them. From now on whenever we need to compare vectors at different points we will parallel transport one of them along this geodesic and suppress~$\gamma$ from the notation.
With respect to the coordinates $(x,y,z) $ on $K$, the non-zero Christoffel symbols read, having set $ x ^1 =x $, $x ^2 =y $, $x ^3 =z $,
\begin{equation*}
\Gamma ^i _{ \ ij }=\Gamma ^i _{ \ ji } =\begin{cases}
 \dfrac{x ^j }{R ^2 - |\mathbf{x} |^2 } & \text{if $j \neq i $}, \vspace{1mm}\\
 \dfrac{2x ^i }{R ^2 - |\mathbf{x} |^2 } & \text{if $j = i $}.
\end{cases}
\end{equation*}
We use rotational symmetry to align the geodesic with the $x$ axis. Then solving~(\ref{pteq}) one finds that the vector at $T _{ \mathbf{x} _2 } K $ obtained by parallel transport of $v$ along $\gamma$ has components $(\tilde v _x, \tilde v _y , \tilde v _z )$ with respect to $(\partial _x , \partial _y , \partial _z ) |_{ \mathbf{x} _2 }$ given by
\begin{gather}
\tilde v _x = \sqrt{ \frac{R ^2 - |\mathbf{x} _2 |^2 }{R ^2 - |\mathbf{x} _1 |^2 }} \left[
v _x + \frac{(x _2 - x _1 ) (\mathbf{x} _1 \cdot \mathbf{v} )}{\mathbf{x} _1 \cdot (\mathbf{x} _2 - \mathbf{x} _1 )} \left( \sqrt{ \frac{R ^2 - |\mathbf{x} _2 |^2 }{R ^2 - |\mathbf{x} _1 |^2 }} -1 \right) \right. \nonumber\\
\left. \hphantom{\tilde v _x = \sqrt{ \frac{R ^2 - |\mathbf{x} _2 |^2 }{R ^2 - |\mathbf{x} _1 |^2 }} \bigg[}{}+ \frac{ (\mathbf{x} _2 \times \mathbf{x} _1 )\cdot \big( (\mathbf{x} _2 - \mathbf{x} _1 )\times \mathbf{v} \big)}{\mathbf{x} _1 \cdot (\mathbf{x} _2 - \mathbf{x} _1)}
\right]
 ,\label{tilvx}\\
\tilde v _y
= \sqrt{ \frac{R ^2 - |\mathbf{x} _2 |^2 }{R ^2 - |\mathbf{x} _1 |^2 }} \left[
v _y + \frac{(y _2 - y _1 ) (\mathbf{x} _1 \cdot \mathbf{v} )}{\mathbf{x} _1 \cdot (\mathbf{x} _2 - \mathbf{x} _1 )} \left( \sqrt{ \frac{R ^2 - |\mathbf{x} _2 |^2 }{R ^2 - |\mathbf{x} _1 |^2 }} -1 \right)
\right] , \nonumber\\
\tilde v _z
= \sqrt{ \frac{R ^2 - |\mathbf{x} _2 |^2 }{R ^2 - |\mathbf{x} _1 |^2 }} \left[
v _z + \frac{(z _2 - z _1 ) (\mathbf{x} _1 \cdot \mathbf{v} )}{\mathbf{x} _1 \cdot (\mathbf{x} _2 - \mathbf{x} _1 )} \left( \sqrt{ \frac{R ^2 - |\mathbf{x} _2 |^2 }{R ^2 - |\mathbf{x} _1 |^2 }} -1 \right) \right] , \nonumber
\end{gather}
where $\cdot $, $\times $ are the dot and cross product of Euclidean 3-space.

Note that if $\mathbf{x} _2 =- \mathbf{x} _1 $ then parallel transport reduces to the identity so that $\tilde{v}_{i}=v_{i}$. Thus, we can compare vectors tangent to antipodal points of $K$ by simply comparing their coordinates just as if we were in flat space. Moreover it can be checked that, denoting by $(P_{ 21 })^a _{ \ b }$ the components of the parallel transport operator with respect to the coordinate frame, so that $ \tilde v ^a =(P_{ 21 }) ^a _{ \ b }v ^b $,
\begin{equation}
\label{ptrid}
\left. \frac{\partial (P _{ 21 })^a _{ \ b }}{\partial x _1 ^i } \right | _{ \mathbf{x} _2 =- \mathbf{x} _1 } =
\left. \frac{\partial (P _{ 21 })^a _{ \ b }}{\partial x _2 ^i } \right | _{ \mathbf{x} _2 =- \mathbf{x} _1 }, \qquad i =1,2,3.
\end{equation}

\subsection{The point particle approximation}
\label{ppa}
A hyperbolic monopole $(\Phi, A )$ on $H ^3 $ is a solution of the Bogomolny equations
\begin{equation}
\label{bogo}
\mathrm{d} _{ A } \Phi = \star F,
\end{equation}
where $\star $ is the Hodge operator with respect to the $H ^3 $ metric. The Bogomolny equations are supplemented by the Prasad--Sommerfeld boundary conditions:
\begin{align}
\label{monp}
p &= \lim _{ r \rightarrow \infty }R\| \Phi \|,\\
\label{monk}
k &= \lim _{ r \rightarrow \infty } \frac{R}{4 \pi p} \int _{ S ^2 _r } \operatorname{Tr} (\Phi F ) \in \mathbb{Z} .
\end{align}
Here $S ^2 _r $ is a 2-sphere of geodesic radius $r$ centred at some fixed point of $H ^3 $, which may conveniently be taken as the origin $r =0 $ of the coordinates used in~(\ref{bgpolargeo}). The value of $p$ is known as the monopole mass, and the integer $k$ is the monopole (magnetic) charge. The framed moduli space $\mathcal{M} _k ^{ \mathrm{hyp}} $ of magnetic monopoles of charge $k$ is the space of solutions of~(\ref{bogo}) satisfying~(\ref{monp}) and~(\ref{monk}) quotiented by the group of framed bundle automorphisms. At least for $2p \in \mathbb{Z} $, the moduli space $\mathcal{M} _k ^{ \mathrm{hyp}} $ is known to be a smooth manifold of dimension $4k $~\cite{atiyah:1988}.

As discussed in Section~\ref{introduction}, the $L ^2 $ metric on $\mathcal{M} _k ^{ \mathrm{hyp}} = \big(S ^1 \times \tilde {\mathcal{M}} _k ^{ \mathrm{hyp}} \big)/\mathbb{Z} _k $ is not well-defined. We now proceed to investigate the dynamics of a point particle approximation to two well-separated monopoles with the aim to understand if this dynamics can be interpreted as geodesic motion with respect to some metric on $\tilde{ \mathcal{M}} _2 ^{ \mathrm{hyp}} $. As we shall see, we are able to do so by restricting to a~4-dimensional submanifold of $\tilde{ \mathcal{M}} _2 ^{ \mathrm{hyp}} $ corresponding to antipodal configurations.

Two well-separated monopoles can be approximated by two point dyons having electric, magnetic and scalar charges. This is a familiar approximation in the case of Euclidean mono\-poles~\mbox{\cite{gibbons:1995, manton:1985}} and has been applied to the study of hyperbolic monopoles in the case where one monopole is moving in the background of several fixed ones~\cite{gibbons:2007}. Here we consider two well-separated monopoles that are both free to move and view them as point particles of equal mass~$m$, with electric and magnetic charges $q_{i}$, $g_{i}$, $i=1,2$, located at the points $\mathbf{x}_{1},\mathbf{x}_{2}\in H ^3 $. As in the Euclidean case, the scalar charge of the $i$-th monopole is $\sqrt{q_{i}^{2}+g_{i}^{2}}$. We will assume that the dyons have the same magnetic charge $g_{1}=g_{2}=g$ and denote by $q$ the difference between the electric charges, $q=q_{2}-q_{1}$.

The 2-particle dynamics can be described in terms of the Lagrangian formalism. The Euclidean case is discussed in~\cite{manton:1985}, which we refer to for the details. The scalar charges modify the rest masses of the particles and the electric charge (respectively magnetic charge) of each particle couples to the Li\'{e}nard--Wiechert 4-potential $ A ^\mu $ (respectively dual 4-potential $\tilde A ^\mu $) produced by the other one. The dual potential $\tilde A ^\mu $ is obtained from $A ^\mu $ via the electromagnetic duality transformation $q _i \rightarrow g _i $, $g _i \rightarrow - q _i $. Keeping terms up to quadratic order in the particle velocities~$v _i $ and the charge difference $q$, in the Euclidean case the resulting Lagrangian is
\begin{equation}
\label{leucl}
L_{\mathrm{E} } = - 2m + \frac{m}{2} \big( |{v} _1 |^2 + |{v} _2 |^2 \big)
+ \frac{\CVe }{8 \pi }
\big( q^2 - g ^2 \vert{v} _2 -{v} _1 \vert^2 \big) + \frac{gq }{4 \pi } \omegae (v _2 -v _1 ),
\end{equation}
where $\CVe = |\mathbf{x} _2 - \mathbf{x} _1 |^{-1} $ and
\begin{equation*}
%\label{omegaeuclidean}
 \omegae =\left(\frac{z_{2}-z_{1}}{\vert \mathbf{x}_{2}-\mathbf{x}_{1}\vert}\right)\left(\frac{(y_{2}-y_{1})\rd x -(x_{2}-x_{1})\rd y}{(x_{2}-x_{1})^{2}+(y_{2}-y_{1})^{2}}\right).
\end{equation*}
If we regard $\CVe $ as a function of $ \mathbf{x} _2 =\mathbf{x} $ only, then $ \rd\omegae = \stare \mathrm{d} \CVe $, where $\stare $ is the Hodge star with respect to the $E ^3 $ metric.

Proceeding in a similar way, we find that the Lagrangian for a 2-particle system in $H ^3 $ is
\begin{gather}
L_{\tu{2P}}=\frac{m}{2} \big( \|v_{1}\|^{2}+\|v_{2}\|^{2}\big) + \frac{\CV}{8\pi}\big(q^{2}-g^{2}\|v_{2}-P_{21}v_{1}\|^{2}\big) \nonumber \\ \hphantom{L_{\tu{2P}}=}{}+\frac{gq}{8\pi}(\omega(v_{2}-P_{21}v_{1})-\omega(v_{1}-P_{12}v_{2})).
\label{l2p}
\end{gather}
Some of the differences between~(\ref{l2p}) and~(\ref{leucl}) simply amount to the replacement of the Euclidean metric with the hyperbolic one:
The Euclidean norm $|\cdot |$ is replaced by the hyperbolic one $\| \cdot \| $ and the Green's function $\CVe $ of the Euclidean Laplacian is replaced by the hyperbolic one $\CV $.
With respect to the coordinates~(\ref{KtoL}) of $K$, $\CV $ is given by
\begin{equation}
\label{scpot}
R \CV= \coth\bigg(\frac{D_K(\mathbf{x}_1, \mathbf{x}_2)}{R}\bigg)-1 ,
\end{equation}
where $D _K $ is hyperbolic distance in the Klein model, see~(\ref{hypdistk}), and the one-form $\omega$ by
\begin{equation}
\label{omega2p}
\omega= \left(\frac{z_{2}-z_{1}}{\vert \mathbf{x}_{2}-\mathbf{x}_{1}\vert}\right)\left(\frac{(y_{2}-y_{1})\rd x -(x_{2}-x_{1})\rd y}{(x_{2}-x_{1})^{2}+(y_{2}-y_{1})^{2}}\right).
\end{equation}
If we consider $\CV$ as a function of $\mathbf{x} _2 =\mathbf{x} $ only, we again have $\mathrm{d} \omega = \star \mathrm{d} \CV $, where $\star $ is now calculated with respect to the hyperbolic metric.

The appearance of the parallel propagator $P _{ 21 }$ is due to the non-zero curvature of $H ^3 $. As previously discussed, it denotes parallel transport along the unique length-minimising geodesic from particle 1 to particle 2 and its expression with respect to the coordinates~(\ref{KtoL}) is given by~(\ref{tilvx}). Since parallel transport is an isometry, $ \|v _2 - P _{ 21 }v _1 \|^2 $ is already invariant under the interchange of particle 1 and 2. However $ (gq/ 4 \pi )  \omega (v _1 - P _{ 21 } v _1 )$ is not invariant and needs to be symmetrised under $1 \leftrightarrow 2 $ as we have done in~(\ref{l2p}) --- recall that $q =q _2 - q _1 $ so $q \rightarrow -q $ under $1 \leftrightarrow 2 $. In the Euclidean case symmetrisation is not needed since parallel transport is trivial.

We now turn to the special case of antipodal configurations, $\mathbf{x} _2 =- \mathbf{x} _1 $. Antipodal configurations of two point dyons correspond to centred ${\rm SU}(2) $ monopoles.
In fact, following~\cite{murray:2003} we take a hyperbolic monopole to be centred if it lies in the zero set of the moment map of the ${\rm SO}(3)\subset {\rm SO}_{0}(1,3)$ action. More intuitively, if we embed the ball model of $H ^3 $ in $\mathbb{R}^{4}$ then a~configuration is centred in the hyperbolic sense if it is centred in the ``Euclidean'' sense. For two monopoles, the latter condition is equivalent to the two monopoles having antipodal centres.

Restricting to antipodal configuration is justified since, as we will now show, dyons starting off at antipodal positions with opposite velocities remain antipodal. In other words, antipodal configurations are preserved by time evolution.

Let $U$ be a vector field generating a symmetry of the Lagrangian $L$, and $\delta _U x ^i _a $ be the infinitesimal change in the Klein--Beltrami coordinates~(\ref{KtoL}) $x ^i _a $ of particle $a$ along $U$. By N\"other's theorem, the conserved quantity associated to $U$ is
\begin{equation*}
C_U = \sum _{ a =1 }^2 \sum _{ i =1 }^3 \frac{\partial L }{\partial v _a ^i } \delta _U x ^i _a .
\end{equation*}
For an interaction potential independent of the particle velocities, N\"other's theorem applied to the symmetries~(\ref{kvfKrot}) gives the conserved quantities
\begin{align}
\label{consangmom}
C _{ X _i } &= \sum _{ a =1 }^2 \left( \frac{ x ^j _a v ^k _a - x _a ^k v ^j _i }{ R ^2 - |\mathbf{x} _a |^2 } \right) ,\\
\label{conslinmom}
 C _{ Y _i }&= \sum _{ a =1 }^2 \left( \frac{ v ^i _a }{ R ^2 - |\mathbf{x} _a |^2 } \right),
\end{align}
where $v ^i _a $ is the $i $-th component of particle $a$ velocity with respect to the coordinate frame $\partial _i $ and~$(ijk )$ is a symmetric permutation of $(123) $. We can recognise~(\ref{consangmom}) and~(\ref{conslinmom}) for the angular and linear momentum, respectively, of the system along the direction $\partial _i $.

Because of the velocity-dependent interactions, the right-hand side of~(\ref{consangmom}) and~(\ref{conslinmom}) has additional terms. However if we differentiate~(\ref{conslinmom}) with the additional terms included, evaluate at antipodal positions $ \mathbf{x} _2 =- \mathbf{x} _1 $ and make use of~(\ref{ptrid}), we still find that the two particles experience opposite accelerations. Thus two particles starting at antipodal positions with opposite velocities will maintain antipodal positions throughout. This choice of initial conditions corresponds to taking the constants $ (C _{ Y _i } )$ to be zero, i.e.,~to zero total linear momentum.

\subsection{The asymptotic moduli space metric}
On the basis of the results of Section~\ref{ppa}, we would like to restrict $L_{\tu{2P}}$ to antipodal configurations.
While spatial inversion $A$ is an isometry of $H ^3 $ and a symmetry of $L_{\tu{2P}}$, the two particles have different electric charges so $A$ is not a symmetry of an antipodal configuration and we cannot invoke the principle of symmetric criticality. However, if $L$ is a 2-particle Lagrangian, and $\La $ is the Lagrangian obtained by setting $ x _2 = F( x _1 ) $ in $L$, it is easy to show that the Euler--Lagrange equations associated to $\La $ are equivalent to those associated to $L$ and restricted to configurations satisfying $x _2 = F (x _1 )$ if and only if $F$ is an affine transformation, i.e.,~$\partial ^2 F / \big(\partial x _1 ^i \big) ^2 =0 $ for all values of $i$. In the present case $F =A = - \operatorname{Id}_3 $.

For ease of notation we give the argument for $i =1 $, the general case is similar. Setting $\La =L\big( x _1 , x _2 = F ( x _1 ) , \dot x _1 , F ^\prime\dot x _1\big) $ the Euler--Lagrange equations associated to $\La $ are
\begin{gather}
\label{aaaarg}
\frac{\mathrm{d} }{\mathrm{d} t} \frac{\partial \La}{\partial \dot x _1 } - \frac{\partial \La}{\partial x_1 }
=\left[ \left( \frac{\mathrm{d} }{\mathrm{d} t} \frac{\partial L}{\partial \dot x _1 }- \frac{\partial L}{\partial x_1 } \right) + \left( \frac{\mathrm{d} }{\mathrm{d} t} \frac{\partial L}{\partial \dot x _2 }- \frac{\partial L}{\partial x_2 } \right) F ^\prime + \frac{\partial L }{\partial \dot x _2 } F ^{ \prime\prime } \dot x _1 \right] _{ x _2 =F (x _1 ) }\!\!\!\!\!\!=0.
\end{gather}
If $F ^{ \prime\prime } = 0 $, then $x _2 = F (x _1 ) = C _1 x_1 + C _2 $, $\partial / \partial x _2 = \frac{1}{C_1} \partial / \partial x _1 $ and~(\ref{aaaarg}) becomes
\begin{equation*}
%\label{aaaarg2}
\frac{\mathrm{d} }{\mathrm{d} t} \frac{\partial \La}{\partial \dot x _1 } - \frac{\partial \La}{\partial x_1 }
=2\left[ \left( \frac{\mathrm{d} }{\mathrm{d} t} \frac{\partial L}{\partial \dot x _1 }- \frac{\partial L}{\partial x_1 } \right) \right] _{ x _2 =F (x _1 ) }=0,
\end{equation*}
showing that the equations associated to $L$ and restricted to $x _2 =F (x _1 )$ are equivalent to those associated to $\La $.

Let us thus consider the Lagrangian $L_{\tu{2P}}$ restricted in such a way. Setting $ x ^i _2 = - x _1 ^i = x ^i $, $P _{ 12 }=P _{ 21 }=1 $, and $v _2 ^i = - v _1 ^i =v ^i$ in~(\ref{l2p}), we obtain
\begin{equation}
\label{reducedlag}
\La=\bigg(m-\frac{g^{2}\CVa}{2\pi}\bigg)\|v\|^{2}+\frac{\CVa}{8\pi}q^{2}+\frac{gq}{2\pi}\omegaa(v),
\end{equation}
where $\CVa $ and $\omegaa $ are the scalar potential and 1-form~(\ref{scpot}) and~(\ref{omega2p}) with $\mathbf{x} =\mathbf{x} _2 =-\mathbf{x} _1 $. It is now convenient to switch to the geodesic polar coordinates of \eqref{bgpolargeo} with $r$ the geodesic distance between $\mathbf{x}$ and $-\mathbf{x}$. Then
\begin{equation*}
R \CVa = \coth \left( \frac{r}{R} \right) -1 , \qquad \omegaa = \cos \theta \, \mathrm{d} \phi ,
\end{equation*}
satisfying
\begin{equation}
\label{abelianmonopole}
\mathrm{d} \omegaa = \star \mathrm{d} \CVa.
\end{equation}
The Lagrangian~(\ref{reducedlag}) is essentially the $k =2 $ case of the Lagrangian obtained in~\cite{gibbons:2007} by considering the motion of one monopole in the background of $k-1$ other fixed ones.
 The analysis to show that the dynamics associated to~(\ref{reducedlag}) can be reinterpreted as geodesic motion now parallels that of~\cite{gibbons:2007} and results in the hyperbolic Taub-NUT (hTN) metric, but for completeness we give the details here. First we add in the constant term $-\frac{m}{4g^{2}}q^{2}$ so that~(\ref{reducedlag}) becomes
\begin{equation}
\La=m\bigg(1-\frac{g^{2}\CVa}{2\pi m}\bigg) \|v\|^{2}-\frac{mq^{2}}{4g^{2}}\bigg(1-\frac{g^{2}\CVa}{2\pi m}\bigg)+\frac{gq}{2\pi}\omegaa(v). \label{eq: 2particle reduced Lagrangian}
\end{equation}
Next we interpret the electric charge as the rate of change of a phase, $q=\dot{\chi}$, and rewrite~(\ref{eq: 2particle reduced Lagrangian}) in the form
\begin{equation}
\lred=m\big[U(r){\|v\|}^{2}+W(r) \ora{R^2 } (\dot{\chi}+\omegaa)^{2}\big].
\label{eq:reduced lagrangian}
\end{equation}
The dynamics associated to~(\ref{eq:reduced lagrangian}) is geodesic motion with respect to the metric
\begin{equation*}
\rd s^{2}=U \gK+W \ora{R ^2 } (\rd\chi+\omegaa )^{2}.
\end{equation*}
The phase $\chi$ is a cyclic variable in \eqref{eq:reduced lagrangian} with conserved momentum
\begin{equation*}
p_{\chi}=2m \ora{R^2}W (\dot{\chi}+\omegaa(v) ) \coloneqq kq,
\end{equation*}
where $k$ is a constant to be determined. Eliminating $\dot{\chi}=\frac{kq}{2m \ora{R^2}W}-\omegaa(v)$ from $\lred$ using the Routhian procedure we obtain
\begin{align}
\lred'&=\lred-p_{\dot{\chi}}\dot{\chi}= m\left[U \|v\|^{2}+W\ora{R ^2 } \left(\frac{kq}{2m\ora{R^2 } W}\right)^{2}\right]-kq\left(\frac{kq}{2m \ora{R ^2 }W}-\omegaa(v)\right)\nonumber\\
&=mU\| v\|^{2}-\frac{k^{2}q^{2}}{4m\ora{R ^2 }W} +kq  \omegaa(v).
\label{afterrouth}
\end{align}
The expression~(\ref{afterrouth}) matches the reduced 2-particle Lagrangian \eqref{eq: 2particle reduced Lagrangian} if
\begin{equation*}
U=\bigg(1-\frac{g^{2}\CVa}{2\pi m}\bigg),\qquad k=\frac{g}{2\pi}, \qquad \frac{1}{W}=\bigg(\frac{2\pi m\ora{R}}{g^{2}}\bigg)^{2}U,
\end{equation*}
so we obtain the metric
\begin{equation}
\rd s^{2}=\bigg(1-\frac{g^{2}\CVa}{2\pi m}\bigg)\gK+\bigg(\frac{g^{2}}{2\pi m\ora{R }}\bigg)^{2}\bigg(1-\frac{g^{2}\CVa}{2\pi m}\bigg)^{-1}\ora{R ^2 }(\rd\chi +\omegaa)^{2}.\label{eq:Hyperbolic Taub-NUT metric}
\end{equation}
Condition~(\ref{abelianmonopole}) implies
$ \rd\omegaa=-\big(2\pi m/g^{2}\big)\star \rd U$.
Working in units where $g^{2}=4\pi m N$, setting
\begin{equation*}
M =- N <0,
\end{equation*}
and introducing the left-invariant 1-form on ${\rm SU}(2) $
\begin{equation*}
\eta _3 = \mathrm{d} \chi + \omegaa =\mathrm{d} \chi + \cos \theta \, \mathrm{d} \phi ,
\end{equation*}
where $\chi $ has range $\chi \in [0, 4 \pi )$ in order to avoid conical singularities, we can rewrite~(\ref{eq:Hyperbolic Taub-NUT metric}) as
\begin{gather}
\ghtn=U\gK+4 M ^{2}U ^{-1} \eta _3 ^{2}, \nonumber\\
U=1 + \frac{ 2 M}{R} \left(\coth \left( \frac{r}{R} \right) -1\right)
=1 + \frac{ 4 M} {R} \big( \mathrm{e} ^{ \frac{2r}{R} } -1\big) ^{-1} ,\label{eq:monopole htn}
\end{gather}
which is the hyperbolic Taub-NUT metric with negative mass $M $. The metric~(\ref{eq:monopole htn}) with positive~$M$ was first introduced in~\cite{lebrun:1991a}, see also~\cite{atiyah:2012a,franchetti:2018,gibbons:2007} for a discussion of its properties.

It is worth pausing to recall some facts about the Taub-NUT metric and its hyperbolic cousin~(\ref{eq:monopole htn}). Both metrics can be expressed in terms of the so-called Gibbons--Hawking ansatz
\begin{equation*}
\mathrm{d} s ^2 = U g _3 + 4 M ^2 U ^{-1} (\mathrm{d} \psi + \alpha )^2 ,
\end{equation*}
where $M$ is a constant, $g _3 $ is either the Euclidean metric on $E ^3 $, for Taub-NUT, or the metric of hyperbolic 3-space $H ^3 $, for hyperbolic Taub-NUT, and the 1-form $\alpha $ satisfies the equation $\mathrm{d} \alpha = \star \mathrm{d} U $, with $\star $ the Hodge star with respect to $g _3 $. As a consequence, $U$ is a Green function of the $g _3 $ Laplacian. In the case of $E ^3 $ by taking
\begin{equation*}
U =1 + \frac{2 M }{r},
\end{equation*}
with $r$ the usual radial coordinate, one obtains the Taub-NUT (TN) metric. If the mass parameter~$M$ is non-negative TN is a smooth complete metric defined on a manifold diffeomorphic to~$\mathbb{R} ^4 $.
In the case of $H ^3 $ by taking $U$ to be the hyperbolic Green's function
\begin{equation*}
U = 1 + \frac{ 4 M} {R} \big( \mathrm{e} ^{ \frac{2r}{R} } -1\big) ^{-1} ,
\end{equation*}
with $r$ the geodesic coordinate of~(\ref{bgpolargeo}), one obtains the hyperbolic Taub-NUT (hTN) metric~(\ref{eq:monopole htn}). As its Euclidean relative, hTN is a smooth complete metric defined on a space diffeomorphic to $\mathbb{R} ^4 $ if $M \geq 0 $ and singular otherwise. The geometry of hTN near the NUT is equal to that of TN and as $R \rightarrow \infty $ the hTN metric with mass $M$ converges to the TN one with the same mass. While in~(\ref{eq:monopole htn}) we have kept the dependence on both the mass parameter~$M$ and the radius of curvature $R$ of $H ^3 $, up to homothety the hTN metric only depends on the ratio $M / R $ as can be checked by substituting $r \rightarrow M r $.

Clearly there are many similarities between TN and hTN. Besides the fact that they both arise from the Gibbons--Hawking ansatz and are defined on diffeomorphic spaces, they both have bi-axial Bianchi IX form, thus admitting a cohomogeneity one action of ${\rm SU}(2) \times {\rm U}(1) $; they both are circle fibrations over a 3-manifold of constant curvature, $E ^3 $ for TN and $H ^3 $ for hTN, except at the NUT $r =0 $, a fixed point of the isometric ${\rm U}(1) $ action where the circle fibre collapses to zero size; they both have an asymptotic circle fibration with fibres of finite length, an asymptotic behaviour called ALF in the Euclidean case. Finally, both TN and hTN admit a multi-(h)TN generalisation with $k$ NUTs obtained by taking $U$ to be the superposition with equal weights of~$k$ poles. There is also a very important difference: while multi-TN is hyperk\"ahler, hyperbolic multi-TN is half-conformally flat but not even Einstein.

\section{Further remarks and conclusions}
\label{sec:further remarks}
The hTN metric may be relevant to the dynamics of hyperbolic monopoles. Besides the results that we have presented here,~\cite{nash:2007} shows how hyperbolic multi-TN with $k$ NUTs emerges as the moduli space of one ${\rm SU}(2) $ monopole with $k$ singularities, a result which also follows from the analysis in~\cite{gibbons:2007} once we reinterpret the fixed monopoles as abelian singularities. The double r\^ole of hyperbolic (multi-)TN with the appropriate value of the mass parameter as both an asymptotic moduli space metric of two centred ${\rm SU}(2) $ monopoles and the moduli space of one~${\rm SU}(2) $ monopole with singularities completely parallels what happens in the Euclidean case. Multi-TN is shown to be the moduli space of singular Euclidean monopoles in~\cite{kronheimer:1985}, where the singular monopoles are-interpreted as smooth circle-invariant instantons on multi TN. The construction in the hyperbolic case is completely similar~\cite{nash:2007} and has been used in~\cite{franchetti:2016} to construct singular as well as smooth hyperbolic monopoles.

It is only natural to ask if the parallelism between the hyperbolic and Euclidean case extends to the full centred 2-monopole space: Is there a complete metric on this moduli space which asymptotically reduces to negative mass hTN? We are now going to show that, at least for a~specific value of the monopole mass, the answer is yes and such a metric is in the conformal class of the family constructed in~\cite{hitchin:1996}.

In~\cite{hitchin:1996}, Hitchin constructed a family $\hito _k $, for $k \geq 3$ an integer, of ${\rm SO}(3) $-invariant metrics defined on the non-compact space $S ^4 \setminus \mathbb{R} P ^2 $. The metric $\hito _k $ is half-conformally-flat and conformally equivalent to an Einstein metric $\hitein_k $ on $S ^4 $ having positive scalar curvature
\begin{equation*}
s _k =2 \tan ^2 \left( \frac{\pi }{k}\right)
\end{equation*}
 with a conical singularity of deficit angle $ \frac{2 \pi }{k-2} $ along an embedded $ \mathbb{R} P ^2 $.
For $k =3 $ there is no conical singularity and $\hitein _3 $ is the round metric on $S ^4 $. For $k =4 $ the metric $\hitein _4 $ admits a smooth global branched cover isometric to $\mathbb{C} P ^2 $ with the Fubini--Study metric. For $k \geq 5 $ the metrics are new.

For our purposes, what matters the most is the fact that for $k \geq 5 $ these metrics are naturally defined on the moduli space of centred ${\rm SU}(2)$ monopoles of charge 2 on $H ^3 $. Taking $H ^3 $ to have curvature $-1 $ the monopoles have mass
\begin{equation*}
p = \frac{k-4 }{4}.
\end{equation*}
Equivalently one could take the monopoles to have unit mass and the curvature of $H ^3 $ to be~$- 1/ p ^2 $. Importantly, as $k \rightarrow \infty $ the scalar curvature $s _k \rightarrow 0 $ and $\hitein _k $ converges to the Ricci-flat Atiyah--Hitchin metric.

 By~\cite{tod:1994}, $\hito _k $ is determined by a solution of Painlev\'e's 6th equation and the conformal factor making $\hito _k $ Einstein can be expressed as an algebraic function of the data determining $\hito _k $. The main problem is thus solving Painlev\'e's equation, which is done in~\cite{hitchin:1996} via twistorial methods. Referring to the original paper for the details, we just note that $\hito _k $ is given by\footnote{Our conventions differ from those used in~\cite{hitchin:1996} by a different normalisation of the left-invariant forms on ${\rm SU}(2) $, $\mathrm{d} \eta _i = - \frac{1}{2} \epsilon _{ ijk }\eta _j \wedge \eta _k $ and relabelling. More precisely $\frac{1}{2} \eta _1 = \sigma _3 $, $ \frac{1}{2}\eta _2 = \sigma _2 $, $\frac{1}{2}\eta_{3} = \sigma _1 $ $\Omega _1 =\Omega _3 $, $ \Omega _3 =\Omega _1 $, where the quantities on the left-hand side (respectively right-hand side) are those used here (respectively in~\cite{hitchin:1996}). }
\begin{equation}
\label{g0}
\hito _k
=	\frac{\mathrm{d} x ^2 }{x (1-x )} + \frac{\eta _3 ^2 }{4\Omega _3 ^2 } + \frac{(1-x ) \eta _2 ^2 }{4\Omega _2 ^2 } + \frac{x  \eta _1 ^2 }{4 \Omega _1 ^2 }.
\end{equation}
The metric $\hito _k $ is negative definite for $x\in (1, \infty )$ and can be extended to $x =1 $, which is a bolt with the topology of $\mathbb{R} P ^2 $.

It is shown in~\cite{hitchin:1996} that for large values of $x$,
\begin{equation}
\label{omegasympt}
\begin{aligned}
\Omega _1 ^2 &\simeq - \frac{(k-2 )^2 }{4k ^2 }, \qquad
\Omega _2 ^2 \simeq \frac{4 ^{ 1-4/k } x ^{ 1-2/k }}{k ^2 },\qquad
\Omega _3 ^2 \simeq -\frac{4 ^{ 1-4/k } x ^{ 1-2/k }}{k ^2 }.
\end{aligned}
\end{equation}
Making the coordinate change
\begin{equation*}
x = 2 ^{ 2 - \frac{k}{2}} u ^{-\frac{k}{2} }
\end{equation*}
in~(\ref{g0}), and
\begin{equation}
\label{htncoordchange}
\frac{2r}{R} = - \log \big( 2 ^{ 3-\frac{12}{k}} u \big)
\end{equation}
in $\ghtn$, see~(\ref{eq:monopole htn}), one finds that near $u =0 $, to leading order,
\begin{equation*}
- \frac{\hito _k }{k ^2 } \simeq \frac{\ghtn}{R ^2 } \simeq \frac{\mathrm{d} u ^2 }{4 u ^2 } + \frac{2 ^{ \frac{12}{k} -5}}{u} \mathrm{d} \Omega ^2 + \frac{\eta _3 ^2 }{(k-2 )^2 },
\end{equation*}
where $\mathrm{d} \Omega ^2 =\eta _1 ^2 + \eta _2 ^2 $ is the round metric on $S ^2 $, provided that the mass parameter takes value
\begin{equation*}
M ^2 = \bigg( \frac{R}{2(k-2 ) } \bigg) ^2 .
\end{equation*}
While $\hito _k $ and $\ghtn$ agree to leading order, the approximation~(\ref{omegasympt}) is not precise enough to determine the sign of $M $.

However the metric $\hito _6 $ is determined exactly in~\cite{hitchin:1996} and given by, for $x=\frac{s^{3}(s+2)}{2s+1}$,
\begin{gather}
\hito _6 = -\frac{36s(1+s)\mathrm{d} s^{2}}{(1+2s)^{2}\big(s^{2}+s-2\big)}-\frac{9s^{2}(s-1)(1+s)^{3}}{(1+2s)\big(1+s+s^{2}\big)^{2}}\eta_{2}^{2}-\frac{9s^{3}(1+s)}{(s-1)(2+s)(1+2s)}\eta_{1}^{2} \nonumber\\
\hphantom{\hito _6 =}{}-\frac{9s^{2}(1+s)}{(s-1)(1+2s)^{2}}\eta_{3}^{2}.
\label{g0k6}
\end{gather}
Making the coordinate change $ s = u ^{-1} $ in~(\ref{g0k6}) and using~(\ref{htncoordchange}) with $k =6 $ in~(\ref{eq:monopole htn}) we now find that for small $u$
\begin{equation*}
- \frac{\hito _6}{36} \simeq \frac{\ghtn}{R ^2 } \simeq
 \frac{\mathrm{d} u ^2 }{4 u ^2 }- \frac{ \mathrm{d} u ^2 }{4u} + \frac{\mathrm{d} \Omega ^2}{8u} + \frac{\eta _3 ^2 }{16 }
\end{equation*}
provided that the hTN mass $M $ takes value
\begin{equation*}
M = - \frac{R}{8}.
\end{equation*}
Thus $\hito _6 $ is asymptotic to hTN with negative mass.

The main point of this work was to show that negative mass hTN emerges as an asymptotic moduli space metric of two centred ${\rm SU}(2) $ monopoles, which was done in Section~\ref{sec:point particles}. This result and the relation between negative mass hTN and $\hito _6 $ which we just discussed invite many further questions which we leave for future work.

First, the asymptotics of $\hito _k $ for general $k$ and its behaviour as $k \rightarrow \infty $ need further study. In particular note that negative mass hTN converges in the zero curvature limit to negative mass~TN, which is the correct asymptotic form of the Atiyah--Hitchin metric, while it is the Einstein metric $\hitein _k $ rather than $\hito _k $ which converges to Atiyah--Hitchin as $k \rightarrow \infty $. It is thus natural to ask what is the $k \rightarrow \infty $ limit of $\hito _k $ and how is it related to the Atiyah--Hitchin metric.

Second, the metric $\hitein _k $ constructed in~\cite{hitchin:1996} is special by virtue of being Einstein, but what makes $\hito _k $ special within its conformal class? At least at the asymptotic level the answer may lie with the Abelian monopole equations satisfied by $(U , \omegaa )$, $\rd\omegaa= \star \rd U$, which are not preserved by a conformal rescaling of the metric. In the Euclidean case the Abelian monopole equations imply that the three self-dual two forms $\omega^{i}=\omega_{E}\wedge \rd x^{i}+\frac{1}{2}\epsilon^{i}_{\;\; jk}\CVe\, \rd x^{j}\wedge \rd x^{k}$ are closed and provide three hyperk\"ahler forms; it is possible that in the hyperbolic case they also determine some special structure, although this remains to be explored.

Many other questions along the lines of ``what is the hyperbolic analogue of'' some property of the Euclidean moduli space metric could be asked. We only mention the following one. Two families of (hyperk\"ahler) gravitational instantons with ALF asymptotics are known: $A _k $, which is the same as multi-TN with $k + 1 $ NUTs, and $D _k $, which includes the Atiyah--Hitchin manifold as $D _0 $. As shown in~\cite{schroers:2021}, ALF $D _k $ manifolds with $k \geq 1 $ can be constructed by gluing NUTs to $D _0 $.
Is it possible to obtain a hyperbolic analogue of ALF $D _k $ by similar means? While the hyperbolic analogues of TN and $D _0 $ are at our disposal, the construction in~\cite{schroers:2021} strongly relies on the hyperk\"ahler structures on $D _0 $ and TN, which are not shared by their hyperbolic relatives.

Finally the case of non-centred configurations of two monopoles is worth investigating. The problem is non-trivial already at the point particle level: While in Euclidean space the possibility of factoring out the centre of mass motion makes the reduced dynamics independent of the total momentum, in the hyperbolic case the dynamics does depend on the total momentum of the system, see~\cite{shchepetilov:2003} and references therein.

\subsection*{Acknowledgements}
GF thanks the Simons Foundation for its support under the Simons Collaboration on Special Holonomy in Geometry, Analysis and Physics [grant number 488631]. CR thanks Michael Singer for useful discussions about the notion of centring for hyperbolic monopoles. The work of~CR was supported by the Engineering and Physical Sciences Research Council [grant number EP/V047698/1].

\pdfbookmark[1]{References}{ref}
\LastPageEnding

\end{document}